\documentclass{Interspeech2024}

\interspeechcameraready

\title{Dirichlet process mixture model based on topologically augmented signal representation for clustering infant vocalizations}

\name[affiliation={1,2}]{Guillem}{Bonafos}
\name[affiliation={3,4}]{Clara}{Bourot}
\name[affiliation={3}]{Pierre}{Pudlo}
\name[affiliation={3}]{Jean-Marc}{Freyermuth}
\name[affiliation={3}]{Laurence}{Reboul}
\name[affiliation={5}]{Samuel}{Tronçon}
\name[affiliation={4}]{Arnaud}{Rey}

\address{
  $^1$UJM, CNRS, Inserm, ENES Lab, St-Étienne, France
  $^2$UJM, CNRS, Hubert Curien Lab, St-Étienne, France
  $^3$Aix Marseille Univ, CNRS, I2M, Marseille, France
  $^4$Aix Marseille Univ, CNRS, LPC, Marseille, France
  $^5$Résurgences R\&D, Arles, France}
\email{}


\DeclareSIUnit\sone{sone}

\begin{document}

\maketitle

\begin{abstract}
    


    Based on audio recordings made once a month during the first 12 months of a child's life, we propose a new method for clustering this set of vocalizations. We use a topologically augmented representation of the vocalizations, employing two persistence diagrams for each vocalization: one computed on the surface of its spectrogram and one on the Takens' embeddings of the vocalization. A synthetic persistent variable is derived for each diagram and added to the MFCCs (Mel-frequency cepstral coefficients). Using this representation, we fit a non-parametric Bayesian mixture model with a Dirichlet process prior to model the number of components. This procedure leads to a novel data-driven categorization of vocal productions. Our findings reveal the presence of 8 clusters of vocalizations, allowing us to compare their temporal distribution and acoustic profiles in the first 12 months of life.

\end{abstract}






















\section{Introduction}

During the first year of life, the vocal productions of human infants undergo a developmental trajectory, actively exploring their acoustic environment through behaviors such as crying, cooing, and babbling. Infants adapt the evolution and diversification of their vocalizations to a target language \cite{terhaarCrossspeciesParallelsBabbling2021} and typically produce their first word by the end of the first year \cite{kuhlEarlyLanguageAcquisition2004}. 
Monitoring these pre-language vocalizations is of great importance, not only for gaining a deeper comprehension of the distinct phases of language development but also for predictive insights into various disorders \cite{ollerPrecursorsSpeechInfancy1999, bartl-pokornyVocalisationRepertoireEnd2022}. 
The use of advanced storage and recording tools allows for the creation of extensive new databases. When combined with innovative statistical analysis techniques, these tools contribute to a deeper exploration of the early stages of language development \cite{ollerAutomatedVocalAnalysis2010, millingSpeechNewBlood2022}.

In this study, we worked with a database that includes vocalizations automatically extracted from long-form audio recordings of a child at home. Recordings were done over her first year, spanning from 0 to 12 months, with three days of recordings per month. The outcome is a longitudinal vocalization database, capturing vocalizations in a real-life setting and diverse contexts.

Our objective is to propose a novel method to categorize vocal productions, without predefining the number of categories, but rather estimating them from the data. To achieve this, we employ a non-parametric Bayesian model, specifically a Dirichlet process mixture model. The clustering process is grounded in a topologically augmented representation of the signal, allowing the incorporation of additional information pertaining to the topology of the vocalizations.

For clustering, we need to represent vocalizations in a low-dimensional space. Topological Data Analysis (TDA), which has demonstrated its efficacy across various domains \cite{salchMathematicsMedicinePractical2021, caoUnsupervisedEnvironmentalSound2019, henselSurveyTopologicalMachine2021}, is a promising candidate for enhancing the current representation of infant vocalizations. Its stability-theoretic properties make it particularly valuable for the examination of natural signals \cite{cohen-steinerStabilityPersistenceDiagrams2007}. The integration of topological information can provide valuable additional information for a more nuanced description of these vocal productions.

In the subsequent sections, we provide an overview of the database in Section~\ref{part:voc_data}. The computation of the augmented topological representation and the clustering model is detailed in Section~\ref{part:modeling}. Section~\ref{part:results_cluster} outlines the clustering results, followed by a comprehensive discussion in Section~\ref{part:discussion}. Finally, we draw conclusions in Section~\ref{part:conclusion_cluster}.

\section{Data}\label{part:voc_data}

The dataset comprises vocalizations of a child, spanning from birth to the child's first birthday. Each vocalization is represented by a stereo-channel audio signal sampled at \SI{44.1}{\kilo\hertz} in PCM format. Extracted from longer audio files, the signals are converted to mono by averaging both channels and then rescaled the pulse modulation signal to a range of $-1$ to $1$.

These vocalizations originate from long-form audio recordings made by the parents of a female French child at home, at regular intervals, over a one-year period. 
Ethical approval was obtained, along with a declaration of conformity for experimental research involving humans, allowing for the recording of human baby vocalizations. 
Parents, equipped with a portable microphone located near the child, recorded audio samples three days a month, capturing various moments throughout the day and night. Following the methodology outlined in \cite{bonafosDetectingHumanNonhuman2023}, we automatically extracted all the segments labeled as baby vocalizations from these continuous recordings, resulting in a dataset of 1924 vocalizations. Unfortunately, due to legal constraints, we are unable to publicly share the data. Vocalizations lasting more than \SI{10}{\second} were excluded, yielding a final set of 1851 vocalizations with an average duration of \SI{2.51}{\second}. 

Table~\ref{tab:effective_voc} provides the distribution of vocalizations detected over the first year. It's noteworthy that vocalizations are not available for every month; specifically, we lack data for the first, fourth, fifth, and tenth months. The absence of vocalizations for these months stems from the inability to conduct recordings during these periods or the absence of detected vocalizations in the recordings. 
\begin{table}[h]
  \caption{Number of vocalizations per month in the long-form audio recordings, as well as the mean and the standard deviation of the duration of the vocalizations produced per month}
  \label{tab:effective_voc}
  \centering
  \begin{tabular}{cccc}
    \toprule
    \textbf{Month} & \textbf{Count} & \textbf{Mean duration} & \textbf{Standard deviation}\\
    \midrule
    $2$ & $667$ & \SI{2.09}{\second} & \SI{1.57}{\second} \\
    $3$ & $139$ & \SI{2.71}{\second} & \SI{2.00}{\second} \\
    $6$ & $132$ & \SI{3.12}{\second} & \SI{2.12}{\second} \\
    $7$ & $154$ & \SI{2.81}{\second} & \SI{1.94}{\second} \\
    $8$ & $159$ & \SI{2.74}{\second} & \SI{2.15}{\second} \\
    $9$ & $285$ & \SI{2.75}{\second} & \SI{2.03}{\second} \\
    $11$ & $212$ & \SI{2.59}{\second} & \SI{1.91}{\second} \\ 
    $12$ & $98$ & \SI{2.62}{\second} & \SI{2.09}{\second} \\
    \bottomrule
  \end{tabular}
\end{table}

\section{Modelling}\label{part:modeling}


\subsection{Topologically augmented signal representation}




We give here the technical details for reproducibility, we refer the reader to \cite{chazalIntroductionTopologicalData2021} for more details. TDA assumes that data has a shape \cite{carlssonTopologyData2009}. We recover this shape through a filtration, a nested sequence of simplicial complexes \cite{zomorodianComputingPersistentHomology2005}. We then derive from the filtration the persistent homology, which serves as a topological descriptor of the data. In our case, we represent each vocalization by using two different objects: the surface of its spectrogram and its Takens' embeddings. Depending on the object, we then adapt the filtration to compute the persistent homology.

For each audio recording, we first compute the spectrogram using a Gaussian window of \SI{11.6}{\milli\second} and a $90\%$ overlap. The spectrogram $S(t,\omega)=|F(t,\omega)|^2$, where $F(t, \omega)$ denotes the Short Time Fourier Transform, defines a surface in $\mathbb{R}^3$, with dimensions representing time $t$, frequency $\omega$ and amplitude $S$. We apply a sublevel set filtration to compute the persistent homology of the spectrogram, \textit{i.e.}, for $f:S \rightarrow \mathbb{R}$, we compute a nested sequence of topological spaces $S_r = f^{-1}(-\infty, r]$ for increasing value of $r$. 

Second, we compute the Takens' embeddings, which embed a time series into a $D$ dimensional Euclidean space using time-delay \cite{takensDetectingStrangeAttractors1981a}. We estimate the time delay parameter $\tau$ such that $\text{AMI}(\tau)<1 / e$, where AMI is the Average Mutual Information. We estimate the embedding dimension $D$ using Cao's algorithm \cite{caoPracticalMethodDetermining1997}. We reduce the dimension $D$ to 3 for all embeddings using UMAP \cite{mcinnesUMAPUniformManifold2020}, ensuring uniformity across all embeddings. This yields the vocalization representation as a point cloud $\mathcal{P}_D = \{p_1,...p_D\}\subset\mathbb{R}^D$ where $p_i=(x_i, x_{i+\tau}, x_{i+2\tau},...,x_{i+(D-1)\tau}).$ We apply an Alpha filtration to compute the persistent homology of the embeddings, involving a nested family of Alpha complex $Alpha(r)=\{\sigma \subseteq \mathcal P | \bigcap_{p \in \sigma}R_x(r) \neq \emptyset\},$ where $R_x(r)$ is the intersection of each Euclidean ball with its corresponding Voronoi cell, for increasing value of $r$.

For both objects, we have an increasing sequence of topological spaces. We compute the homology at all scales, \textit{i.e.}, for all $r$ of the sequence. We resume in a persistence diagram the persistence homology of the object, where a point in a diagram has two coordinates, the value $r$ of its birth and the value $r$ of its death. Persistent homology then yields a multiscale topological description of the object \cite{wassermanTopologicalDataAnalysis2018}.

Persistence diagrams cannot be used directly for statistical analysis. We therefore extract information from the diagrams by computing a set of variables: persistent entropy \cite{atienzaPersistentEntropySeparating2019}, $p$-norm of the diagram \cite{cohen-steinerLipschitzFunctionsHave2010}, persistent Betti number \cite{edelsbrunnerComputationalTopologyIntroduction2009}, and descriptors of the vector collecting the lifetime of the points of the diagram following \cite{fireaizenAlarmSoundDetection2022, pereiraPersistentHomologyTime2015}. From this set of variables for each diagram, we compute a synthetic persistent variable using PCA. The first principal component of the PCA is retained, explaining $27.79\%$ of the variance of the set of variables from the persistence diagram of the spectrogram surface, and $65.94\%$ of the variance of the set of variables from the persistence diagram of Takens' embeddings.

In addition to the topological features, we compute Mel Frequency Cepstral Coefficients (MFCC), classical frequency descriptors of human speech analysis \cite{sueurSoundAnalysisSynthesis2018}. We compute twelve coefficients, with a window length of \SI{25}{\milli\second} and an overlap of $40$\%. We take the average of the twelve coefficients to ensure a consistent number of MFCC for all vocalizations.

The resulting topologically augmented representation of vocalizations comprises fourteen dimensions: 12 MFCC and 2 synthetic persistent variables, with one summarizing the persistence diagram computed on the surface of the spectrogram and another summarizing the persistence diagram computed on Takens' embeddings.

\subsection{Nonparametric Bayesian modelling for clustering}

We aim to determine the number of clusters in the dataset $X$ using a Dirichlet process mixture model. The mixture model is defined as $p(\bm x) = \int_{\Theta} f(\bm x; \theta) d G(\theta)$, where $\Theta$ is the parameter space, $f$ is a $p$-dimensional Gaussian kernel. Consequently, $\theta = (\bm \mu, \bm \Sigma)$, and $\Theta = \mathbb R^p \times S_+^p$, where $S_+^p$ is the space of semi-definite positive $p\times p$ matrices. We are interested in determining the number $K$ of mixture components and the assignment of vocalizations to these components. To learn the complexity of the model on the data, we set a Dirichlet process as a prior on $G$ and define the model as follows:
\begin{equation}\label{eq:hierarchical_model}
    \begin{split}
        \bm x_i | \theta_i &\sim f(\bm x_i; \theta_i) \qquad i=1,\ldots,N \\
        \theta_i | G &\sim G \\
        G &\sim DP(\alpha, G_0),
    \end{split}
\end{equation}
where $DP$ is a Dirichlet process. See \cite{Teh2010a} for further reading.
We choose the concentration parameter $\alpha$ such that $\mathbb E[K|n,\alpha] = \sum_{i=1}^n \frac{\alpha}{\alpha + i - 1}$ \cite{dorazioSelectingPriorPrecision2009} to favor values around $K=5$, following expectations based on prior knowledge \cite{cychoszBabbleCorCrosslinguisticCorpus2019}.

For conjugacy, we put a normal-inverse Wishart prior on the base measure $G_0$, with $\Sigma_j \sim IW(\nu_0, \Sigma_0)$ and $\mu_j | \Sigma_j \sim \mathcal N (\bm m_0, \Sigma_0 / k_0)$. We use non-informative priors for the degree of freedom of the inverse Wishart, $\nu_0=p$, and hyperpriors $\bm m_0 \sim \mathcal N(\bm m_1, \bm S_1)$, $k_0 \sim \mathcal Gamma (\tau_1, \xi_1)$, and $\Sigma_0 \sim W(\nu_1, \bm \Sigma_1)$. 
We follow an empirical Bayes procedure to calibrate the hyperparameters on the dataset:
$\bm m_1$ is the mean of each dimension of $X$, $\bm S_1$ is the variance-covariance matrix, $\tau_1 = \xi_1 = 1$, $\nu_1 = p + 2$, and $\bm \Sigma_1 = \bm S_1 / 2$.

We utilize the collapsed Gibbs sampler of \cite{nealMarkovChainSampling2000}, based on the Chinese Restaurant Process representation, to sample the indicator variable $z = \{z_i\}_{i=1}^N$, which assigns each vocalization to a latent cluster by marginalizing mixture weights and parameters. This assignment gives us the clustering, and we run the MCMC with 10,000 iterations, discarding the first 4,000 as burn-in.
As a Bayesian model, the posterior provides a distribution on possible clusterings rather than a single point estimate. 
Following \cite{wadeBayesianClusterAnalysis2018}, we select the best clustering by specifying a loss function of the true clustering. The loss function used is the Variation of Information and the estimate is the one that minimizes the posterior expected loss.

\subsection{Acoustics differences between clusters}

After obtaining the partition, we proceed to compare the different clusters by computing various acoustic descriptors. Subsequently, we utilize these descriptors as input for a multinomial logit model, where the cluster serves as the response variable. We estimate one model per cluster, treating each cluster as the referential group.
For each of the eight models, we assess the statistical differences in each acoustic descriptor between the referential group and the other clusters. 
This analysis provides insights into the distinctive acoustic characteristics associated with each cluster, helping to characterize and differentiate them based on the selected features.

\section{Data analysis}\label{part:results_cluster}

\subsection{Partition}

Our model identifies 8 distinct clusters. Initially, we detected 9 clusters, but one of them comprised only 5 records, and none of these records included baby vocalizations. This cluster essentially served as a "garbage cluster," grouping false positives that remained in the dataset. The Dirichlet process mixture model, leveraging our topologically augmented representation, effectively groups together recordings that differ from the rest of the dataset. It automatically recognizes and segregates a "garbage" class, helping eliminate false positives.

Table~\ref{tab:prop_month_cluster} provides a summary of the cluster distribution by month of the year, indicating the proportion of production for each cluster during each month. This breakdown offers insights into how vocalization patterns vary across different clusters and months.

\begin{table}[h]
  \caption{Proportion (percentage) of production for each cluster during the year}
  \label{tab:prop_month_cluster}
  \centering
  \begin{tabular}{ccccccccc}
    \toprule
    & \multicolumn{8}{c}{\textbf{Month}}\\
    \cmidrule{2-9}
    \textbf{Cluster} & 2 & 3 & 6 & 7 & 8 & 9 & 11 & 12 \\
    \midrule
    \textbf{1} & 15 & 16 & 2 & 9 & 11 & 22 & 19 & 6 \\
    \textbf{2} & 11 & 4 & 1 & 7 & 16 & 33 & 20 & 8 \\
    \textbf{3} & 65 & 2 & 2 & 2 & 7 & 15 & 4 & 3 \\
    \textbf{4} & 31 & 23 & 0 & 8 & 0 & 31 & 8 & 0 \\
    \textbf{5} & 83 & 1 & 4 & 2 & 3 & 1 & 6 & 1 \\
    \textbf{6} & 24 & 11 & 4 & 7 & 16 & 4 & 24 & 9 \\
    \textbf{7} & 22 & 10 & 20 & 16 & 8 & 6 & 10 & 8 \\
    \textbf{8} & 0 & 2 & 3 & 15 & 2 & 73 & 3 & 3 \\
    \bottomrule
    \end{tabular}
\end{table}

\subsection{Comparison of clusters}

The insights from Table~\ref{tab:prop_month_cluster} reveal distinctive temporal patterns among the clusters. First, Cluster 2 is characterized by late vocalizations in the first year, with a substantial portion (a third) produced in the ninth month and a notable increase in production during the eighth month. About 20\% of its production occurs during the eleventh month, indicating that more than half of the cluster's production takes place after the ninth month. Similar to Cluster 2, Cluster 8 represents late vocalizations, primarily produced from the ninth month onwards.
In contrast to Clusters 2 and 8, Cluster 5 comprises vocalizations produced predominantly in the first months of life, with over 80\% occurring during this period. This cluster encapsulates early vocalizations, which decrease as the child learns to produce other types of vocalizations. Like Cluster 5, Cluster 3 also contains vocalizations primarily produced in the first two months, constituting 65\% of its production.
Clusters 4 and 7 exhibit a skewed distribution towards the first few months of life, with the majority of vocalizations produced in the first 6 months, and even just the first three months for Cluster 4. However, there is comparatively more vocal production from these clusters over the rest of the year than Clusters 3 and 5.
Clusters 1 and 6 stand out for being produced throughout the entire year, indicating a more consistent vocalization pattern across the different months.

\begin{figure}
    \centering
    \includegraphics[width=\linewidth]{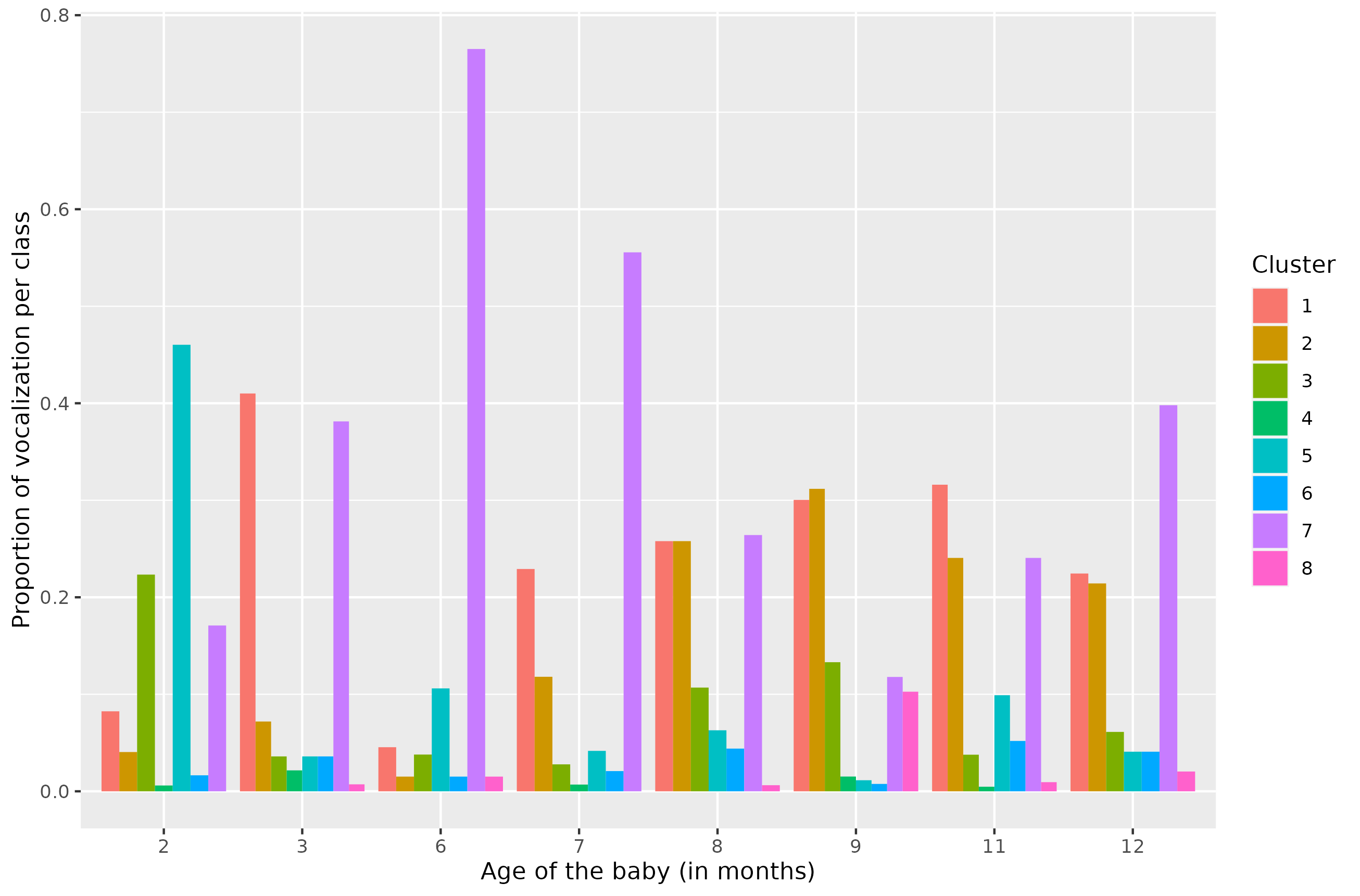}
    \caption{Proportion of monthly production of vocalization per cluster. Parents did not record during three months, yet the gap.}
    \label{fig:prop_month_cluster}
\end{figure}

These temporal variations in vocalization patterns highlight the diversity of the clusters and the developmental changes in vocal behavior over the course of the first year of life, that we illustrate in Figure~\ref{fig:prop_month_cluster}. Whereas we have the proportion of vocalization of each cluster per month (\textit{i.e.}, it sums to one per cluster) in Table~\ref{tab:prop_month_cluster}, we plot in Figure~\ref{fig:prop_month_cluster} the proportion of vocalization of each cluster at each month (\textit{i.e.}, it sums to one per month). 

We present a summary of median acoustic descriptors per cluster in Table~\ref{tab:acoustics}. Utilizing these descriptors, we employed multinomial logit models to estimate acoustic differences among the identified clusters. To enhance clarity, we provide a concise overview of the main results and distinctions between clusters. Specifically, we highlight instances where the parameter associated with a descriptor is statistically different from zero for the majority of other levels (\textit{i.e.}, from other clusters).

Clusters 2 and 8, characterized as late vocalizations, exhibit differences from other clusters. Cluster 2 differs in its proportion of voiced frames, entropy level, and $F_3$. Cluster 8, on the other hand, varies in spectral centroid level, entropy level, $F_2$, and $F_3$.

Clusters 3 and 5, representing earlier vocalizations, also show distinctions from other clusters. Cluster 3 differs in its proportion of voiced frames, spectral centroid level, loudness, and $F_3$. Cluster 5 exhibits differences in its proportion of voiced frames, spectral centroid level, entropy level, Harmonics-to-Noise Ratio, loudness, Frequency Modulation, and $F_3$.

Cluster 4 stands out from others due to differences in its proportion of voiced frames and entropy level. Cluster 7 differs in the proportion of voiced frames, spectral centroid level, loudness, $F_2$, and $F_3$. Cluster 1 exhibits distinctions in the proportion of voiced frames, spectral centroid level, Frequency Modulation, loudness, roughness, $F_1$, and $F_3$. Lastly, Cluster 6 varies in its proportion of voiced frames, spectral centroid level, and $F_3$.

\begin{table*}[th]
  \caption{Acoustics descriptors of each cluster. We report the median for each cluster. Unit of Pitch, Formants, Spectral Centroid and FM is \unit{\hertz}, Loudness is in \unit{\sone}, HNR is in \unit{\dB}, Duration is in \unit{\sec}, Voiced and Roughness are proportions.}
  \label{tab:acoustics}
  \centering
  \begin{tabular}{ccccccccccccc}
    \toprule
    \textbf{Cluster} & \textbf{Duration} & \textbf{Pitch} & \textbf{F\textsubscript{1}} & \textbf{F\textsubscript{2}} & \textbf{F\textsubscript{3}} & \textbf{Voiced} & \textbf{SC} & \textbf{Entropy} & \textbf{HNR} & \textbf{FM} & \textbf{Loudness} & \textbf{Roughness}\\
    \midrule
    \textbf{1} & $1.4$ & $301$ & $725$ & $2374$ & $4640$ & $46$ & $3084$ & $0.33$ & $9.57$ & $6.23$ & $12.4$ & $22.7$ \\
    \textbf{2} & $2.6$ & $334$ & $832$ & $2041$ & $3757$ & $48$ & $2596$ & $0.24$ & $12.5$ & $5.46$ & $13.0$ & $20.5$ \\
    \textbf{3} & $1.6$ & $338$ & $1534$ & $3571$ & $6000$ & $17$ & $2268$ & $0.27$ & $10.7$ & $5.46$ & $7.15$ & $23.1$ \\
    \textbf{4} & $6.6$ & $342$ & $878$ & $2610$ & $4804$ & $16$ & $1870$ & $0.35$ & $10.5$ & $5.00$ & $11.2$ & $23.3$ \\
    \textbf{5} & $1.4$ & $316$ & $913$ & $3056$ & $5685$ & $30$ & $3769$ & $0.41$ & $8.41$ & $6.16$ & $10.3$ & $23.8$ \\
    \textbf{6} & $2.8$ & $382$ & $925$ & $2546$ & $4244$ & $43$& $3549$ & $0.36$ & $8.23$ & $5.46$ & $13.7$ & $22.6$ \\
    \textbf{7} & $2.2$ & $346$ & $856$ & $2117$ & $3724$ & $28$ & $3252$ & $0.39$ & $10.4$ & $5.44$ & $17.1$ & $21.2$ \\ 
    \textbf{8} & $1.6$ & $381$ & $1034$ & $2384$ & $4065$ & $23$ & $8528$ & $0.48$ & $9.29$ & $5.46$ & $12.8$ & $23.8$ \\
    \bottomrule
  \end{tabular}
\end{table*}

\section{Discussion}\label{part:discussion}

We conducted an analysis on a unique dataset comprising 1851 vocalizations from a baby, spanning her birth to her first birthday. These vocalizations were extracted from longitudinal recordings captured at home, devoid of external interactions. Employing an innovative topologically augmented signal representation, we adapted an unsupervised strategy to cluster the vocalizations based on this representation.

Remarkably, certain clusters of vocalizations only emerge after a specific period, while others exhibit a decreasing production trend. Clusters 3 and 5, prominently generated post-birth, experience minimal production throughout the remainder of the year. It is plausible to hypothesize that these clusters represent the initial vocalization classes, serving as a foundation for subsequent vocalization categories.

In contrast, clusters 2 and 8 materialize towards the end of the year, predominantly around the ninth month, with cluster 8 showing distinctive formant characteristics. These late-emerging clusters coincide with the child's increased diversification of vocal productions and heightened babbling, a phenomenon documented in the literature, peaking between the ninth and tenth months \cite{ollerFunctionalFlexibilityInfant2013, cychoszVocalDevelopmentLarge2021}.

From a language development perspective, aligning with the concept of calibration \cite{terhaarCrossspeciesParallelsBabbling2021}, the child undergoes a learning process during the initial months, gradually mastering her vocal apparatus to produce sounds resembling the phonemes of her native language. Notably, an early vocalization cluster such as cluster 5 also exhibits nonlinear phenomena, indicative of strong vocal tension \cite{koutseff_acoustic_2018}.

Cluster 5, primarily produced in the initial two months, stands out with a notably high entropy level. On the contrary, Cluster 2, produced later in the year, distinguishes itself from other clusters with lower entropy. This suggests an improvement in the child's motor control of the buco-phonatory apparatus over the course of the year, resulting in vocalizations with lower entropy compared to earlier productions. The identified clusters, with diverse temporal distributions, highlight variations in precocity among vocal productions.

The incorporation of topological information in signal representation proves effective in clustering vocalizations based on various acoustic parameters. Notably, we observe no differences in pitch between clusters. Given that we have only one child in this database and that pitch is a good individual marker \cite{lockhart-bouronInfantCriesConvey2023}, our model does not rely on this feature for clustering. 

However, our approach exhibits limitations. First, our model treats all vocalizations as exchangeable, neglecting time dependence. To do this, we need to consider the temporal aspect by employing a non-parametric regression \cite{quintanaDependentDirichletProcess2022}, enabling the incorporation of covariates.

Moreover, while the current topological representation aids in identifying clusters with distinct acoustic profiles, refinement is needed. Synthetic persistent variables are constructed to mitigate the curse of dimensionality, resulting in information loss, particularly for the persistent homology of the spectrogram. Exploring methods for constructing a lower-dimensional signal representation that incorporates topological information, such as \cite{moorTopologicalAutoencoders2020, trofimovLearningTopologypreservingData2023} is a worthwhile avenue.

In terms of modeling, incorporating expert knowledge and refining priors, especially the choice of $\alpha$, could enhance the clustering process. Adjusting $\alpha$ might impact the final clustering outcome, and its initial computation based on expecting five clusters could be refined based on the diversity of vocalizations observed in literature during a baby's first year \cite{ollerFunctionalFlexibilityInfant2013, jhangEmergenceFunctionalFlexibility2017}.

Finally, the current analysis, although valuable for its longitudinal and ecological nature, pertains to a single child. The results, while insightful, cannot propose a new categorization of vocal productions. Future research should deepen the analysis by including more children. Introducing hierarchy in subsequent analyses could facilitate comparisons of vocal productions and their evolution over the first year of life, considering integrated covariates.


\section{Conclusion}\label{part:conclusion_cluster}

In conclusion, we investigated a novel database comprising vocalizations extracted from long-form audio recordings of a child from birth to her first birthday. This dataset offers a unique perspective, capturing vocalizations in an uncontrolled, longitudinal setting without interaction, allowing for the exploration of new inquiries.

We introduced an innovative approach to analyze this database, aiming to identify distinct clusters of vocalizations produced by the child. Employing an unsupervised methodology, we utilized a Dirichlet process mixture model without specifying the number of classes beforehand. By incorporating topological information into the signal representation, we successfully identified 8 vocalization classes throughout the year.

Acknowledging the outlined limitations, the detected clusters exhibited varying production proportions over time. Furthermore, our topologically augmented representation facilitated the identification of clusters with diverse acoustic profiles, illustrating the child's evolving motor control of her buco-phonatory apparatus.

\bibliographystyle{IEEEtran}
\bibliography{biblio}

\begin{thebibliography}{10}
\providecommand{\url}[1]{#1}
\csname url@samestyle\endcsname
\providecommand{\newblock}{\relax}
\providecommand{\bibinfo}[2]{#2}
\providecommand{\BIBentrySTDinterwordspacing}{\spaceskip=0pt\relax}
\providecommand{\BIBentryALTinterwordstretchfactor}{4}
\providecommand{\BIBentryALTinterwordspacing}{\spaceskip=\fontdimen2\font plus
\BIBentryALTinterwordstretchfactor\fontdimen3\font minus \fontdimen4\font\relax}
\providecommand{\BIBforeignlanguage}[2]{{%
\expandafter\ifx\csname l@#1\endcsname\relax
\typeout{** WARNING: IEEEtran.bst: No hyphenation pattern has been}%
\typeout{** loaded for the language `#1'. Using the pattern for}%
\typeout{** the default language instead.}%
\else
\language=\csname l@#1\endcsname
\fi
#2}}
\providecommand{\BIBdecl}{\relax}
\BIBdecl

\bibitem{terhaarCrossspeciesParallelsBabbling2021}
S.~M. {ter Haar}, A.~A. Fernandez, M.~Gratier, M.~Kn{\"o}rnschild, C.~Levelt, R.~K. Moore, M.~Vellema, X.~Wang, and D.~K. Oller, ``Cross-species parallels in babbling: Animals and algorithms,'' \emph{Philosophical Transactions of the Royal Society B: Biological Sciences}, vol. 376, no. 1836, p. 20200239, Oct. 2021.

\bibitem{kuhlEarlyLanguageAcquisition2004}
P.~K. Kuhl, ``Early language acquisition: Cracking the speech code,'' \emph{Nature Reviews Neuroscience}, vol.~5, no.~11, pp. 831--843, Nov. 2004.

\bibitem{ollerPrecursorsSpeechInfancy1999}
D.~K. Oller, R.~E. Eilers, A.~R. Neal, and H.~K. Schwartz, ``Precursors to speech in infancy: {{The}} prediction of speech and language disorders,'' \emph{Journal of Communication Disorders}, vol.~32, no.~4, pp. 223--245, Jul. 1999.

\bibitem{bartl-pokornyVocalisationRepertoireEnd2022}
K.~D. {Bartl-Pokorny}, F.~B. Pokorny, D.~Garrido, B.~W. Schuller, D.~Zhang, and P.~B. Marschik, ``Vocalisation {{Repertoire}} at the {{End}} of the {{First Year}} of {{Life}}: {{An Exploratory Comparison}} of {{Rett Syndrome}} and {{Typical Development}},'' \emph{Journal of Developmental and Physical Disabilities}, Mar. 2022.

\bibitem{ollerAutomatedVocalAnalysis2010}
D.~K. Oller, P.~Niyogi, S.~Gray, J.~A. Richards, J.~Gilkerson, D.~Xu, U.~Yapanel, and S.~F. Warren, ``Automated vocal analysis of naturalistic recordings from children with autism, language delay, and typical development,'' \emph{Proceedings of the National Academy of Sciences of the United States of America}, vol. 107, no.~30, pp. 13\,354--13\,359, Jul. 2010.

\bibitem{millingSpeechNewBlood2022}
M.~Milling, F.~B. Pokorny, K.~D. {Bartl-Pokorny}, and B.~W. Schuller, ``Is {{Speech}} the {{New Blood}}? {{Recent Progress}} in {{AI-Based Disease Detection From Audio}} in a {{Nutshell}},'' \emph{Frontiers in Digital Health}, vol.~4, p. 886615, May 2022.

\bibitem{salchMathematicsMedicinePractical2021}
A.~Salch, A.~Regalski, H.~Abdallah, R.~Suryadevara, M.~J. Catanzaro, and V.~A. Diwadkar, ``From mathematics to medicine: {{A}} practical primer on topological data analysis ({{TDA}}) and the development of related analytic tools for the functional discovery of latent structure in {{fMRI}} data,'' \emph{PLOS ONE}, vol.~16, no.~8, p. e0255859, Aug. 2021.

\bibitem{caoUnsupervisedEnvironmentalSound2019}
Y.~Cao, S.~Zhang, F.~Yan, W.~Li, F.~Sun, and H.~Sun, ``Unsupervised {{Environmental Sound Classification Based On Topological Persistence}},'' in \emph{2019 {{IEEE International Conference}} on {{Signal}}, {{Information}} and {{Data Processing}} ({{ICSIDP}})}, Dec. 2019, pp. 1--5.

\bibitem{henselSurveyTopologicalMachine2021}
F.~Hensel, M.~Moor, and B.~Rieck, ``A {{Survey}} of {{Topological Machine Learning Methods}},'' \emph{Frontiers in Artificial Intelligence}, vol.~4, p. 681108, May 2021.

\bibitem{cohen-steinerStabilityPersistenceDiagrams2007}
D.~{Cohen-Steiner}, H.~Edelsbrunner, and J.~Harer, ``Stability of {{Persistence Diagrams}},'' \emph{Discrete \& Computational Geometry}, vol.~37, no.~1, pp. 103--120, Jan. 2007.

\bibitem{bonafosDetectingHumanNonhuman2023}
G.~Bonafos, P.~Pudlo, J.-M. Freyermuth, T.~Legou, J.~Fagot, S.~Tron{\c c}on, and A.~Rey, ``Detecting human and non-human vocal productions in large scale audio recordings,'' Feb. 2023.

\bibitem{chazalIntroductionTopologicalData2021}
F.~Chazal and B.~Michel, ``An {{Introduction}} to {{Topological Data Analysis}}: {{Fundamental}} and {{Practical Aspects}} for {{Data Scientists}},'' \emph{Frontiers in Artificial Intelligence}, vol.~4, 2021.

\bibitem{carlssonTopologyData2009}
G.~Carlsson, ``Topology and data,'' \emph{Bulletin of the American Mathematical Society}, vol.~46, no.~2, pp. 255--308, 2009.

\bibitem{zomorodianComputingPersistentHomology2005}
A.~Zomorodian and G.~Carlsson, ``Computing {{Persistent Homology}},'' \emph{Discrete \& Computational Geometry}, vol.~33, no.~2, pp. 249--274, Feb. 2005.

\bibitem{takensDetectingStrangeAttractors1981a}
F.~Takens, ``Detecting strange attractors in turbulence,'' in \emph{Dynamical {{Systems}} and {{Turbulence}}, {{Warwick}} 1980}, D.~Rand and L.-S. Young, Eds.\hskip 1em plus 0.5em minus 0.4em\relax {Berlin, Heidelberg}: {Springer Berlin Heidelberg}, 1981, vol. 898, pp. 366--381.

\bibitem{caoPracticalMethodDetermining1997}
L.~Cao, ``Practical method for determining the minimum embedding dimension of a scalar time series,'' \emph{Physica D: Nonlinear Phenomena}, vol. 110, no.~1, pp. 43--50, Dec. 1997.

\bibitem{mcinnesUMAPUniformManifold2020}
L.~McInnes, J.~Healy, and J.~Melville, ``{{UMAP}}: {{Uniform Manifold Approximation}} and {{Projection}} for {{Dimension Reduction}},'' Sep. 2020.

\bibitem{wassermanTopologicalDataAnalysis2018}
L.~Wasserman, ``Topological {{Data Analysis}},'' \emph{Annual Review of Statistics and Its Application}, vol.~5, no.~1, pp. 501--532, 2018.

\bibitem{atienzaPersistentEntropySeparating2019}
N.~Atienza, R.~{Gonzalez-Diaz}, and M.~Rucco, ``Persistent entropy for separating topological features from noise in vietoris-rips complexes,'' \emph{Journal of Intelligent Information Systems}, vol.~52, no.~3, pp. 637--655, Jun. 2019.

\bibitem{cohen-steinerLipschitzFunctionsHave2010}
D.~{Cohen-Steiner}, H.~Edelsbrunner, J.~Harer, and Y.~Mileyko, ``Lipschitz {{Functions Have L}} p -{{Stable Persistence}},'' \emph{Foundations of Computational Mathematics}, vol.~10, no.~2, pp. 127--139, Apr. 2010.

\bibitem{edelsbrunnerComputationalTopologyIntroduction2009}
H.~Edelsbrunner and J.~Harer, \emph{Computational {{Topology}}: {{An Introduction}}}.\hskip 1em plus 0.5em minus 0.4em\relax {AMS Press}, 2009.

\bibitem{fireaizenAlarmSoundDetection2022}
T.~Fireaizen, S.~Ron, and O.~Bobrowski, ``Alarm {{Sound Detection Using Topological Signal Processing}},'' in \emph{{{ICASSP}} 2022 - 2022 {{IEEE International Conference}} on {{Acoustics}}, {{Speech}} and {{Signal Processing}} ({{ICASSP}})}.\hskip 1em plus 0.5em minus 0.4em\relax {Singapore, Singapore}: {IEEE}, May 2022, pp. 211--215.

\bibitem{pereiraPersistentHomologyTime2015}
C.~M. Pereira and R.~F. {de Mello}, ``Persistent homology for time series and spatial data clustering,'' \emph{Expert Systems with Applications}, vol.~42, no. 15-16, pp. 6026--6038, Sep. 2015.

\bibitem{sueurSoundAnalysisSynthesis2018}
J.~Sueur, \emph{Sound Analysis and Synthesis with {{R}}}, 1st~ed.\hskip 1em plus 0.5em minus 0.4em\relax {New York, NY}: {Springer Berlin Heidelberg}, 2018.

\bibitem{Teh2010a}
Y.~W. Teh, ``{D}irichlet processes,'' in \emph{Encyclopedia of Machine Learning}.\hskip 1em plus 0.5em minus 0.4em\relax Springer, 2010.

\bibitem{dorazioSelectingPriorPrecision2009}
R.~M. Dorazio, ``On selecting a prior for the precision parameter of {{Dirichlet}} process mixture models,'' \emph{Journal of Statistical Planning and Inference}, vol. 139, no.~9, pp. 3384--3390, Sep. 2009.

\bibitem{cychoszBabbleCorCrosslinguisticCorpus2019}
M.~Cychosz, A.~Seidl, E.~Bergelson, M.~Casillas, G.~Baudet, A.~S. Warlaumont, C.~Scaff, L.~Yankowitz, and A.~Cristia, ``{{BabbleCor}}: {{A Crosslinguistic Corpus}} of {{Babble Development}} in {{Five Languages}},'' Oct. 2019.

\bibitem{nealMarkovChainSampling2000}
R.~M. Neal, ``Markov {{Chain Sampling Methods}} for {{Dirichlet Process Mixture Models}},'' \emph{Journal of Computational and Graphical Statistics}, vol.~9, no.~2, pp. 249--265, 2000.

\bibitem{wadeBayesianClusterAnalysis2018}
S.~Wade and Z.~Ghahramani, ``Bayesian {{Cluster Analysis}}: {{Point Estimation}} and {{Credible Balls}} (with {{Discussion}}),'' \emph{Bayesian Analysis}, vol.~13, no.~2, pp. 559--626, Jun. 2018.

\bibitem{ollerFunctionalFlexibilityInfant2013}
D.~K. Oller, E.~H. Buder, H.~L. Ramsdell, A.~S. Warlaumont, L.~Chorna, and R.~Bakeman, ``Functional flexibility of infant vocalization and the emergence of language,'' \emph{Proceedings of the National Academy of Sciences}, vol. 110, no.~16, pp. 6318--6323, Apr. 2013.

\bibitem{cychoszVocalDevelopmentLarge2021}
M.~Cychosz, A.~Cristia, E.~Bergelson, M.~Casillas, G.~Baudet, A.~S. Warlaumont, C.~Scaff, L.~Yankowitz, and A.~Seidl, ``Vocal development in a large-scale crosslinguistic corpus,'' \emph{Developmental Science}, vol.~24, no.~5, Sep. 2021.

\bibitem{koutseff_acoustic_2018}
\BIBentryALTinterwordspacing
A.~Koutseff, D.~Reby, O.~Martin, F.~Levrero, H.~Patural, and N.~Mathevon, ``\BIBforeignlanguage{en}{The acoustic space of pain: cries as indicators of distress recovering dynamics in pre-verbal infants},'' \emph{\BIBforeignlanguage{en}{Bioacoustics}}, vol.~27, no.~4, pp. 313--325, Oct. 2018. [Online]. Available: \url{https://www.tandfonline.com/doi/full/10.1080/09524622.2017.1344931}
\BIBentrySTDinterwordspacing

\bibitem{lockhart-bouronInfantCriesConvey2023}
M.~{Lockhart-Bouron}, A.~Anikin, K.~Pisanski, S.~Corvin, C.~Cornec, L.~Papet, F.~Levr{\'e}ro, C.~Fauchon, H.~Patural, D.~Reby, and N.~Mathevon, ``Infant cries convey both stable and dynamic information about age and identity,'' \emph{Communications Psychology}, vol.~1, no.~1, pp. 1--15, Oct. 2023.

\bibitem{quintanaDependentDirichletProcess2022}
F.~A. Quintana, P.~M{\"u}ller, A.~Jara, and S.~N. MacEachern, ``The {{Dependent Dirichlet Process}} and {{Related Models}},'' \emph{Statistical Science}, vol.~37, no.~1, pp. 24--41, Feb. 2022.

\bibitem{moorTopologicalAutoencoders2020}
M.~Moor, M.~Horn, B.~Rieck, and K.~Borgwardt, ``Topological {{Autoencoders}},'' in \emph{{{arXiv}}:1906.00722 [Cs, Math, Stat]}, vol. PMLR 119, 2020, pp. 7045--7054.

\bibitem{trofimovLearningTopologypreservingData2023}
I.~Trofimov, D.~Cherniavskii, E.~Tulchinskii, N.~Balabin, E.~Burnaev, and S.~Barannikov, ``Learning topology-preserving data representations,'' in \emph{The {{Eleventh International Conference}} on {{Learning Representations}}}, Feb. 2023.

\bibitem{jhangEmergenceFunctionalFlexibility2017}
Y.~Jhang, ``Emergence of {{Functional Flexibility}} in {{Infant Vocalizations}} of the {{First}} 3 {{Months}},'' \emph{Frontiers in Psychology}, vol.~8, p.~11, 2017.

\end{thebibliography}

\end{document}